\documentclass[12pt]{article}
\usepackage{amsmath,amsfonts,amssymb}  
\usepackage{graphicx}

 \ifx\pdfoutput\undefined
      \relax
 \else
      \usepackage{epstopdf}
 \fi

\makeatletter
\def\numberbysection{\@addtoreset{equation}{section}
    \def\theequation{\thesection.\arabic{equation}}}
\makeatother

\numberbysection

\newcommand{\be}{\begin{eqnarray}}
\newcommand{\ee}{\end{eqnarray}}
\newcommand{\non}{\nonumber}

\newcommand{\sh}{\mathop{\rm sh}\nolimits}
\newcommand{\ch}{\mathop{\rm ch}\nolimits}

\newcommand{\bff}{\ensuremath{\mathsf{b}}}

\newcommand{\Gf}{\ensuremath{\mathsf{G}}}
\newcommand{\Pf}{\ensuremath{\mathsf{P}}}
\newcommand{\yf}{\ensuremath{\mathsf{y}}}

\def\chi{\chi}

\def\La{\Lambda}

\newcommand{\beq}{\begin{equation}}
\newcommand{\eeq}{\end{equation}}
\newcommand{\bea}{\begin{eqnarray*}}
\newcommand{\eea}{\end{eqnarray*}}
\newcommand{\beqa}{\begin{eqnarray}}
\newcommand{\eeqa}{\end{eqnarray}}

\begin{document}

\begin{titlepage}
\strut\hfill
\vspace{.5in}
\begin{center}

\LARGE A note on the IR limit of the NLIEs\\
of boundary supersymmetric sine-Gordon model\\ [1.0in]
\large Rajan Murgan\footnote{email: rmurgan@svsu.edu}\\[0.8in]
\large Department of Physics,\\ 
\large Saginaw Valley State University,\\ 
\large 7400 Bay Road University Center,  MI 48710 USA\\
      
\end{center}

\vspace{.5in}

\begin{abstract}
We consider the infrared (IR) limit of the nonlinear integral equations (NLIEs) for the boundary supersymmetric 
sine-Gordon (BSSG) model, previously obtained from the NLIEs for the inhomogeneous 
open spin-$1$ XXZ quantum spin chain with general integrable boundary terms,  for values of the boundary 
parameters which satisfy a certain constraint. In particular, we compute the boundary $S$ matrix and determine the
``lattice - IR'' relation for the BSSG parameters.

\end{abstract}

\end{titlepage}

\setcounter{footnote}{0}

\section{Introduction}\label{sec:intro}
One-dimensional spin systems such as the open XXZ quantum spin chain, e.g., see \cite{Ga}-\cite{CRS} and references herein, have been crucial in the 
study of ground and excited states of the corresponding field theoretical models i.e., the boundary sine-Gordon (BSG) \cite{GZ} and boundary supersymmetric 
sine-Gordon (BSSG) models \cite{IOZ, NepSSG}. In particular, studies involving the crucial ultraviolet (UV) limit and infrared (IR) limit which include the computation 
of boundary $S$ matrices (reflection factors) for soliton and breathers, have attracted much interests over the years. One effective tool for such studies is the method of 
nonlinear integral equations (NLIEs). The NLIEs have been obtained for SG model from the Bethe ansatz equations of spin-$1/2$ XXZ spin chain. These 
NLIEs have revealed important results about these models. The studies initiated for periodic boundary conditions \cite {DdV}-\cite{DdV3} were also extended to Dirichlet
and more general boundary conditions \cite{FS}-\cite{ABNPT}. They have been used to compute quantities such as the $S$ matrices (both bulk and boundary), 
bulk and boundary energies, central charges and conformal dimensions. 

The supersymmetric counterpart, the SSG model \cite{ssg1}-\cite{Murgan3}, have also received equally extensive attention. Initially, a set of NLIEs for the close 
spin-$1$ XXZ chain, which is related to the SSG model, were derived in \cite{Su1, Su} from the $T$-$Q$ equations of close spin-$1$ XXZ spin 
chain, by exploiting the analyticity of the transfer matrix eigenvalues described by these equations. Such a method have been later used to derive the desired NLIEs 
describing SSG models with periodic boundary conditions \cite{ArpadRavSuz}, the Dirichlet \cite {ANS} and also for a more general boundary condition \cite{Murgan3}. 
These NLIEs were used to calculate finite-size properties of ground and excited states. 

Our motivation for the present work comes from these efforts. Utilizing solutions for the open spin-$1$ XXZ chain with general integrable boundary terms (rather than 
the diagonal boundary terms used in \cite{ANS}) \cite{IOZ2} that have been recently proposed \cite{FNR}\footnote{The lattice boundary parameters satisfy 
a pair of real constraint.}, a corresponding set of NLIEs was derived for a BSSG model \cite{Murgan3}. We thus plan to analyze the IR limit of these NLIEs along the line 
of \cite{ANS} and obtain the corresponding boundary $S$ matrix. We will show that the result agrees with that obtained by Bajnok {\it et al.} in \cite{BPT}. 
We shall consider only the one-hole state in this paper. Another important objective of this work is to determine the relation between the boundary ``lattice-IR'' parameters 
for the BSSG case \footnote{Note that in \cite{BPT}, two BSSG models are discussed: BSSG${}^{-}$ and BSSG${}^{+}$. Our results are related to the latter model 
by Bajnok {\it et al.}} (for other than that of the Dirichlet boundary condition case). Such a relation is well-known for the BSG case, 
and has been found for BSSG models with Dirichlet boundary conditions \cite{ANS}, labelled there as the Dirichlet BSSG${}^{+}$ model.  

The outline of the article is as follows. In section 2,  we briefly review the open spin-$1$ XXZ quantum spin chain, namely the Hamiltonian, the $T$-$Q$ equations 
and the NLIEs of the model. The BSSG model and the set of NLIEs that describe the model are reviewed in section 3, which are reproduced 
from \cite{Murgan3}. In section 4, we give the main results of the paper. We consider the IR limit of the NLIEs for a state with one hole which is subsequently 
used to compute the boundary $S$ matrix. We then determine the relation between the boundary ``lattice-IR'' parameters.  Finally, a brief discussion of our results and 
some open problems conclude the paper in section 5.

\section{Open spin-$1$ XXZ chain}\label{sec:XXZ}
Since the set of NLIEs that describe BSSG models and the NLIEs for the inhomogeneous open spin-$1$ XXZ quantum spin chains are closely related, in this section, 
we briefly review some crucial results on the open spin-$1$ XXZ quantum spin chain which include the Hamiltonian of the open spin-$1$ XXZ chain, 
the $T$-$Q$ equations for the inhomogeneous model, important auxiliary functions and the NLIEs. The readers are urged to refer to \cite{FNR, Murgan3} for more details on these 
materials. 

\subsection{Hamiltonian and $T$-$Q$ equations}
We begin with the Hamiltonian of the open spin-$1$ XXZ chain. We adopt the notations used in \cite{FNR},
\be
{\cal H} = \sum_{n=1}^{N-1}H_{n,n+1} + H_{b} \,,
\label{Hamiltonianspin1}
\ee
where $H_{n,n+1}$ represents the bulk terms. Explicitly, these terms are given by \cite{ZF},
\be 
H_{n,n+1} &=&  \sigma_{n} - (\sigma_{n})^{2}
+ 2 \sh^2 \eta \left[ \sigma_{n}^{z} + (S^z_n)^2
+ (S^z_{n+1})^2 - (\sigma_{n}^{z})^2 \right] \non \\
&-& 4 \sh^2 (\frac{\eta}{2})  \left( \sigma_{n}^{\bot} \sigma_{n}^{z}
+ \sigma_{n}^{z} \sigma_{n}^{\bot} \right) \,, \label{bulkhamiltonianspin1}
\ee 
where
\be
\sigma_{n} = \vec S_n \cdot \vec S_{n+1} \,, \quad
\sigma_{n}^{\bot} = S^x_n S^x_{n+1} + S^y_n S^y_{n+1}  \,, \quad
\sigma_{n}^{z} = S^z_n S^z_{n+1} \,, 
\ee 
and $\vec S$ are the $su(2)$ spin-1 generators. $H_{b}$ represents the boundary terms which have the following form (see e.g., \cite{IOZ2})
\be 
H_{b} &=& a_{1} (S^{z}_{1})^{2}  + a_{2} S^{z}_{1} 
+  a_{3} (S^{+}_{1})^{2}  +  a_{4} (S^{-}_{1})^{2}  +
a_{5} S^{+}_{1}\, S^{z}_{1}  + a_{6}  S^{z}_{1}\, S^{-}_{1} \non \\
&+& a_{7}  S^{z}_{1}\, S^{+}_{1} + a_{8} S^{-}_{1}\, S^{z}_{1} 
+ (a_{j} \leftrightarrow b_{j} \mbox{ and } 1 \leftrightarrow N) \,,
\ee
where $S^{\pm} = S^{x} \pm i S^{y}$. The coefficients $\{ a_{i} \}$ 
of the boundary terms at site 1 are functions  
of the boundary parameters ($\alpha_{-}, \beta_{-},
\theta_{-}$) and the bulk anisotropy parameter $\eta$. They are given by,
\be
a_{1} &=& \frac{1}{4} a_{0} \left(\ch 2\alpha_{-} - \ch 
2\beta_{-}+\ch \eta \right) \sh 2\eta 
\sh \eta \,,\non \\
a_{2} &=& \frac{1}{4} a_{0} \sh 2\alpha_{-} \sh 2\beta_{-} \sh 2\eta \,, \non \\
a_{3} &=& -\frac{1}{8} a_{0} e^{2\theta_{-}} \sh 2\eta 
\sh \eta \,, \non \\
a_{4} &=& -\frac{1}{8} a_{0} e^{-2\theta_{-}} \sh 2\eta 
\sh \eta \,, \non \\
a_{5} &=&  a_{0} e^{\theta_{-}} \left(
\ch \beta_{-}\sh \alpha_{-} \ch {\eta\over 2} +
\ch \alpha_{-}\sh \beta_{-} \sh {\eta\over 2} \right)
\sh \eta \ch^{\frac{3}{2}}\eta \,, \non \\
a_{6} &=&  a_{0} e^{-\theta_{-}} \left(
\ch \beta_{-}\sh \alpha_{-} \ch {\eta\over 2} +
\ch \alpha_{-}\sh \beta_{-} \sh {\eta\over 2} \right)
\sh \eta \ch^{\frac{3}{2}}\eta \,, \non \\
a_{7} &=&  -a_{0} e^{\theta_{-}} \left(
\ch \beta_{-}\sh \alpha_{-} \ch {\eta\over 2} -
\ch \alpha_{-}\sh \beta_{-} \sh {\eta\over 2} \right)
\sh \eta \ch^{\frac{3}{2}}\eta \,, \non \\
a_{8} &=&  -a_{0} e^{-\theta_{-}} \left(
\ch \beta_{-}\sh \alpha_{-} \ch {\eta\over 2} -
\ch \alpha_{-}\sh \beta_{-} \sh {\eta\over 2} \right)
\sh \eta \ch^{\frac{3}{2}}\eta \,,
\ee
where 
\be
a_{0}= \left[
\sh(\alpha_{-}-{\eta\over 2})\sh(\alpha_{-}+{\eta\over 2})
\ch(\beta_{-}-{\eta\over 2})\ch(\beta_{-}+{\eta\over 2})\right]^{-1} 
\,.
\ee
Similarly, the coefficients $\{ b_{i} \}$ of the boundary terms at 
site $N$ which are functions of  
the boundary parameters ($\alpha_{+}, \beta_{+}, \theta_{+}$) and $\eta$, are given by the following correspondence,
\be
b_{i} = a_{i}\Big\vert_{\alpha_{-}\rightarrow \alpha_{+}, 
\beta_{-}\rightarrow -\beta_{+}, \theta_{-}\rightarrow \theta_{+}} \,.
\ee
We stress beforehand that the NLIEs in \cite{Murgan3} were derived from the $T$-$Q$ equations for the (inhomogeneous) open spin-$1$ XXZ chain with integrable 
boundary terms as given in \cite{FNR}, namely when the boundary parameters $\big(\alpha_{\pm}\,, \beta_{\pm}\,, \theta_{\pm}\big)$
obey the following constraint \cite{Nep2, Nep3},\cite{FNR},\footnote{More generally, the last term in (\ref{constraint}) is $k\eta$, where $k$ is an odd integer. 
Following \cite{Murgan3}, we take $k=1$. See the reference for details. Also refer to \cite{FNR} for more complete discussion on this.}
\be
\alpha_{-} + \beta_{-} + \alpha_{+} + \beta_{+} = \pm (\theta_{-} - 
\theta_{+}) + \eta \,.
\label{constraint}
\ee
As adopted in \cite{AN}, a convenient redefinition of bulk and boundary parameters as follows
\be
\eta = i \mu \,, \qquad 
\alpha_{\pm} = i \mu a_{\pm} \,, \qquad \beta_{\pm} = \mu b_{\pm} \,,
\qquad \theta_{\pm} = i \mu c_{\pm}
\label{newparams} \,,
\ee 
results in the following pair of real constraints: 
\be
a_{-} + a_{+} &=& \pm |c_{-} - c_{+}| + 1 \,, \non \\
b_{-} + b_{+} &=& 0 \,.
\label{realconstraints}
\ee
where $\mu\,, a_{\pm}\,, b_{\pm}\,, c_{\pm}$ are all real, with $0 < \mu < \frac{\pi}{2}$.

Next, we review the model's $T$-$Q$ equations that describe the eigenvalues of the commuting transfer matrices \cite{FNR}. Two relevant commuting transfer matrices for 
spin-$1$ XXZ chain are $T_1(u)$ with a spin-$\frac{1}{2}$ (two-dimensional) auxiliary space, and $T_2(u)$ with a spin-1 (three-dimensional) auxiliary space. 
The eigenvalues of $T_{1}(u)$,  $\Lambda_1(u)$, can be written as 
(following \cite{ANS} and adopting the notations used there)
\beqa
\Lambda_1(u)&=&l_1(u)+l_2(u) \,, \non\\
l_1(u)&=&\sinh(2u+i\mu) \tilde{B}^{(+)}(u) \phi(u+i\mu) {Q(u-i\mu)\over Q(u)} 
\,, \non \\
l_2(u)&=&\sinh(2u-i\mu) \tilde{B}^{(-)}(u) \phi(u-i\mu) {Q(u+i\mu)\over Q(u)} 
\,, \label{T1} 
\eeqa
where 
\be
\phi(u)&=&\sinh^N(u-\La) \sinh^N(u+\La)\,, \non \\
\tilde{B}^{(\pm)}(u)&=&\sinh(u\pm \frac{i\mu A_{+}}{2}) \sinh(u\pm \frac{i\mu A_{-}}{2})\cosh(u\mp \frac{i\mu B_{+}}{2}) \cosh(u\mp \frac{i\mu B_{-}}{2}) 
\,,\non \\
Q(u)&=&\prod_{k=1}^{M}\sinh(u-\tilde{v}_k) \sinh(u+\tilde{v}_k) \,. 
\label{phiBpm}  
\ee
As in \cite{Murgan3}, the lattice boundary parameters have been redefined as $A_{\pm} = 2a_{\pm} - 1\,, B_{\pm} = 2ib_{\pm} + 1$. $\La$ is the inhomogeneity 
parameter, $N$ is the number of spins and $M = N$ represents the number of Bethe roots which are also the zeros $\tilde{v}_k$ of $Q(u)$. 
For this particular model of the XXZ chain, one generally needs to consider two groups of transfer matrix eigenvalues,
labelled as $\tilde{\Lambda}^{(\frac{1}{2},1)(\pm)}(u)$ in \cite{FNR}, to obtain all $3^{N}$ energy eigenvalues. However, we restrict 
to only one of them that contains the ground state, namely $\tilde{\Lambda}^{(\frac{1}{2},1)(-)}(u)$. Refer to \cite{FNR} for greater details on this.
From the fusion relation, the eigenvalues of $T_{2}(u)$,  $\Lambda_2(u)$ can be written as (see e.g. \cite{FNR})
\be
\Lambda_2(u) = \Lambda_1(u-{i\mu\over{2}})\, \Lambda_1(u+{i\mu\over{2}}) - f(u)
\label{T1T1}
\ee
where 
\be
f(u) &=& \phi(u+\frac{3i\mu}{2})\phi(u-\frac{3i\mu}{2})\sinh(2u-2i\mu)\sinh(2u+2i\mu)\tilde{B}^{(+)}(u+\frac{i\mu}{2})\tilde{B}^{(-)}(u-\frac{i\mu}{2})\non\\
&=&  l_1(u+\frac{i\mu}{2})l_2(u-\frac{i\mu}{2})
\label{f}
\ee
Using (\ref{f}), (\ref{T1T1}) can thus be written as 
\be
\Lambda_2(u) &=& l_2(u-\frac{i\mu}{2})l_2(u+\frac{i\mu}{2})+l_1(u-\frac{i\mu}{2})l_2(u+\frac{i\mu}{2})+l_1(u-\frac{i\mu}{2})l_1(u+\frac{i\mu}{2})\non \\
&=& \sinh (2u)\tilde{\Lambda}_2(u)
\label{T1T2}
\ee
where 
\be
\tilde{\Lambda}_2(u) &=& \sinh(2u-2i\mu) \tilde{B}^{(-)}(u-\frac{i\mu}{2}) \tilde{B}^{(-)}(u+\frac{i\mu}{2})\phi(u-\frac{3i\mu}{2}) \phi(u-\frac{i\mu}{2}){Q(u+\frac{3i\mu}{2})\over Q(u-\frac{i\mu}{2})} \non\\
&+&\sinh(2u) \tilde{B}^{(+)}(u-\frac{i\mu}{2}) \tilde{B}^{(-)}(u+\frac{i\mu}{2})\phi(u-\frac{i\mu}{2}) \phi(u+\frac{i\mu}{2}){Q(u+\frac{3i\mu}{2})\over Q(u-\frac{i\mu}{2})} {Q(u-\frac{3i\mu}{2})\over Q(u+\frac{i\mu}{2})}\non\\
&+&\sinh(2u+2i\mu) \tilde{B}^{(+)}(u+\frac{i\mu}{2}) \tilde{B}^{(+)}(u-\frac{i\mu}{2})\phi(u+\frac{3i\mu}{2}) \phi(u+\frac{i\mu}{2}){Q(u-\frac{3i\mu}{2})\over Q(u+\frac{i\mu}{2})} \non\\
&=& \tilde{\lambda}_{1}(u) + \tilde{\lambda}_{2}(u) + \tilde{\lambda}_{3}(u)
\label{Ttilde}
\ee
From (\ref{T1T2}) and (\ref{Ttilde}), one has the following
\be
\tilde{\lambda}_{1}(u) = \frac{l_2(u-\frac{i\mu}{2})l_2(u+\frac{i\mu}{2})}{\sinh (2u)}\,,\quad \tilde{\lambda}_{2}(u) = \frac{l_1(u-\frac{i\mu}{2})l_2(u+\frac{i\mu}{2})}{\sinh (2u)}\,,\quad \tilde{\lambda}_{3}(u) = \frac{l_1(u-\frac{i\mu}{2})l_1(u+\frac{i\mu}{2})}{\sinh (2u)}
\non\\
\label{def}
\ee
The next crucial review are the following auxiliary functions $b(u)$ and $\bar{b}(u)$ that enter the NLIEs:
\be
b(u) = {\tilde{\lambda}_{1}(u) + \tilde{\lambda}_{2}(u) \over \tilde{\lambda}_{3}(u)}\,, \qquad \bar{b}(u) = b(-u) = {\tilde{\lambda}_{3}(u) + \tilde{\lambda}_{2}(u) \over \tilde{\lambda}_{1}(u)}
\label{b}
\ee
For real $u$, $\bar b(u)$ is the complex conjugate of $b(u)$. 
Using (\ref{T1}) and (\ref{def}), $b(u)$ becomes
\be
b(u) ={\Lambda_1(u-{i\mu\over{2}})\over{\sinh(2u+2i\mu)}}{\phi(u-{i\mu\over{2}})\over{
\phi(u+{i\mu\over{2}})\phi(u+{3i\mu\over{2}})}}
{\tilde{B}^{(-)}(u+{i\mu\over{2}})\over{\tilde{B}^{(+)}(u-{i\mu\over{2}})\tilde{B}^{(+)}(u+{i\mu\over{2}})}}
{Q(u+{3i\mu\over{2}})\over{Q(u-{3i\mu\over{2}})}}
\label{bb}
\ee
Finally, we conclude this section with another important auxiliary function, $y(u)$ that enters the NLIEs,
\be
y(u) = \frac{\sinh (2u)\tilde{\Lambda}_{2}(u)}{f(u)}\,,
\label{y}
\ee
\subsection{NLIEs for inhomogeneous open spin-$1$ XXZ chain}\label{sec:NLIEXXZ}

Next, the NLIEs are reviewed as presented in \cite{Murgan3}. It is sufficient here to mention that the
NLIEs are obtained utilizing the analyticity of $\ln \check{\Lambda}_2(u) = \ln \frac{\tilde{\Lambda}_{2}(u)}{\kappa (u)}$ near the real axis, namely,
\beq
0=\oint_C du\ [\ln\check{\Lambda}_2(u)]'' e^{iku}
\label{cauchy1}
\eeq
where the contour $C$ is chosen as in figure below, $\epsilon$ is small and positive. Refer to analysis in \cite{Murgan3} for details.

\setlength{\unitlength}{0.8cm}
\begin{picture}(6,4)(-9,-3)
\put(-2.5,0){\line(1,0){5}}
\put(0,-1.5){\line(0,1){3}}
\put(-1.0,0.3){\vector(-1,0){0.2}}
\put(-2.0,0.3){\line(1,0){4}}
\put(1.0,-0.3){\vector(1,0){0.2}}
\put(-2.0,-0.3){\line(1,0){4}}
\put(0.75,0.5){$\footnotesize{\textrm{$C_{1}$}}$}
\put(0.75,-0.8){$\footnotesize{\textrm{$C_{2}$}}$}
\put(2.0,-0.3){\line(0,1){0.6}}
\put(-2.0,-0.3){\line(0,1){0.6}}
\put(2.2,0.1){$\footnotesize{\textrm{$i\epsilon$}}$}
\put(2.2,-0.3){$\footnotesize{\textrm{$-i\epsilon$}}$}
\put(-2.0,-2.2){$\small{\textrm{Integration\, contour}}$}
\end{picture}

\noindent $\kappa (u)$ is any function whose only real root is a simple zero at the origin, that is $\kappa (0) = 0\,, \kappa '(0) \neq 0$. The prime
denotes differentiation with respect to $u$. These NLIEs are written in coordinate space:
\be
\ln b(u) &=& 
\int_{-\infty}^{\infty}du'\ G(u-u'-i\epsilon) 
\ln (1 + b(u'+i\epsilon)) 
-\int_{-\infty}^{\infty}du'\ G(u-u'+i\epsilon) \ln (1+ \bar 
b(u'-i\epsilon))\non \\
&+& \int_{-\infty}^{\infty}du'\ G_{2}(u-u'+i\epsilon) \ln 
(1+ y(u'-i\epsilon)) + i 2N \tan^{-1}\left({\sinh \frac{\pi u}{\mu}\over 
\cosh \frac {\pi \La}{\mu} } \right) \non \\
&+& i\, \int_{0}^{u}du'\ R(u')  - i\pi\,, 
\non \\
\ln \bar b(u) &=& 
-\int_{-\infty}^{\infty}du'\ G(u-u'-i\epsilon) 
\ln (1 + b(u'+i\epsilon)) 
+\int_{-\infty}^{\infty}du'\ G(u-u'+i\epsilon) \ln (1 + \bar 
b(u'-i\epsilon))\non \\
&+& \int_{-\infty}^{\infty}du'\ G_{2}(u'-u+i\epsilon) \ln 
(1 + y(u'+i\epsilon)) - i 2N \tan^{-1}\left({\sinh \frac{\pi u}{\mu}\over 
\cosh \frac {\pi \La}{\mu} } \right) \non \\
&-& i\, \int_{0}^{u}du'\ R(u')  + i\pi\,, \non \\
\ln y(u) &=& \int_{-\infty}^{\infty}du'\ 
G_{2}(u-u'+i\epsilon) \ln (1 + \bar 
 b(u'-i\epsilon)) + \int_{-\infty}^{\infty}du'\ 
G_{2}(u'-u+i\epsilon) \ln (1 + 
b(u'+i\epsilon))  \non \\
&+& 4\pi i\, \int_{0}^{u}du'\ G_{2}(-u')  \,,
\label{NLIEspin1u}
\ee
where $G(u)$ and $G_{2}(u)$ are the Fourier transforms 
of $\hat G(k)$ and $\hat G_{2}(k)$ respectively \footnote{ 
Conventions used are
\be
\hat f(k) \equiv \int_{-\infty}^\infty e^{i k u}\ 
f(u)\ du \,, \qquad\qquad
f(u) = {1\over 2\pi} \int_{-\infty}^\infty e^{-i k u}\ 
\hat f(k)\ dk \,. \non
\ee}. These are given below,
\be
\widehat G(k)&=&
{\sinh\left((\pi-3\mu)\frac{k}{2}\right)\over
2\cosh{\mu k\over{2}}\sinh\left((\pi-2\mu)\frac{k}{2}\right)} \,, 
\label{Gspin1} \\
\widehat G_{2}(k)&=&{e^{-{\mu k\over{2}}}\over{e^{{\mu 
k\over{2}}}+e^{-{\mu k\over{2}}}}}\,,
\label{G2def}
\ee
Similarly, $R(u)$ refers to the Fourier transform of $\hat R(k)$ which is given below,
\be
\hat R(k) &=& 2\pi \Bigg\{  
{s_{+}\sinh\left((\pi-\mu |A_{+}|)\frac{k}{2}\right)
   +s_{-}\sinh\left((\pi-\mu |A_{-}|)\frac{k}{2}\right)
\over 2\cosh({\mu k\over 2}) \sinh\left((\pi-2\mu)\frac{k}{2}\right)} 
\non \\
& & + {\sinh\left(\frac{k}{2}\mu B_{+}\right)
   +\sinh\left(\frac{k}{2}\mu B_{-}\right)
\over 2\cosh({\mu k\over 2}) \sinh\left((\pi-2\mu)\frac{k}{2}\right)} + {\cosh({\mu k\over 4}) \sinh\left((3\mu-\pi)\frac{k}{4}\right)\over
  \cosh({\mu k\over 2}) \sinh\left((2\mu-\pi)\frac{k}{4}\right)} 
\Bigg\} \,,
\label{Rk}
\ee
where $s_{\pm}\equiv$ sgn($A_{\pm}$) or in terms of $a_{\pm}$ is sgn($2a_{\pm}-1$).

\section{The BSSG model and the NLIEs}\label{sec: bSSG}
In this section, we briefly review the BSSG model, mainly reproduced from \cite{NepSSG}. In addition, we also give the corresponding NLIEs.

\subsection{The BSSG model} 

The Euclidean-space action of the
BSSG model is given by
\be
S &=& \int_{-\infty}^{\infty} dy \int_{-\infty}^{0} dx\ {\cal L}_{0}
+ \int_{-\infty}^{\infty} dy\  {\cal L}_{b} \,,
\ee
where the bulk Lagrangian density is given by
\be
{\cal L}_{0} =
2 \partial_{z}\varphi \partial_{\bar z} \varphi  
- 2 \bar \psi \partial_{z} \bar \psi 
+ 2 \psi \partial_{\bar z} \psi 
- 4 \cos \varphi - 4 \bar \psi \psi \cos {\varphi\over 2} \,,
\label{bulkL}
\ee
In (\ref{bulkL}), $\psi$ and $\bar \psi$ are the two components of a Majorana
Fermion field, and $z=x+iy$, $\bar z=x-iy$.  The boundary Lagrangian at $x=0$ is given by \footnote{This corresponds to the BSSG${}^{+}$ models in \cite{BPT}, 
which we are interested in here.}
\be
{\cal L}_{b} = \bar \psi \psi + ia \partial_{y} a 
- 2 p(\varphi) a (\psi - \bar \psi) + {\cal B}(\varphi) \,,
\label{boundL}
\ee
where $a$ is a Hermitian Fermionic boundary degree of freedom. The functions $p(\varphi)$ and ${\cal B}(\varphi)$, which are
potentials that are functions of the scalar field $\varphi$, are determined from the requirement of boundary integrability and supersymmetry. They are found to be 
\cite{NepSSG}
\be
{\cal B}(\varphi) = 2 \upsilon \cos {1\over 2}(\varphi - \varphi_{0}) \,, \qquad p(\varphi) = {\sqrt{F}\over 2}\sin{1\over 4}(\varphi - D)\qquad 
\mbox{where} \quad 
\tan {D\over 2} ={\upsilon \sin{\varphi_{0}\over 2}\over
\upsilon \cos{\varphi_{0}\over 2} -4} \,.
\label{Bpotential}
\ee
where $F = \sqrt{\upsilon^{2}-8\upsilon\cos \frac{\varphi_{0}}{2}+16}$

The parameters $\upsilon$ and $\varphi_{0}$ are arbitrary and real.
In the limit that the boundary mass parameters tend to infinity, one arrives at the BSSG models with Dirichlet boundary conditions studied in \cite{ANS} (labelled as 
Dirichlet BSSG${}^{+}$ there). Such a BSSG model corresponds to the open spin-$1$ XXZ chain with diagonal boundary terms \cite{MNRitt}.

\subsection{NLIEs for BSSG model}\label{sec:NLIESSG}

Our final review is the set of NLIEs that describes the BSSG model. In the continuum limit, which consists of taking $\La
\rightarrow \infty$, $N \rightarrow \infty$ and lattice spacing $\Delta
\rightarrow 0$, such that the interval length $L$
and the soliton mass $m$ are given by 
\be
L = N \Delta \,, \qquad m={2\over \Delta} e^{-{\pi \Lambda\over \mu}} \,,
\label{continuumlimit}
\ee
respectively, (as stated in earlier works \cite{ArpadRavSuz, ANS}) the NLIEs of the inhomogeneous spin-$1$ XXZ chains lead to the 
NLIEs of the SSG models. With the following change of variable 
\be 
\theta = \frac{\pi u}{\mu} \,.
\label{renormrapidity}
\ee 
(\ref{NLIEspin1u}) becomes
\be
\ln \bff(\theta) &=& 
\int_{-\infty}^{\infty}d\theta'\ \Gf(\theta-\theta'-i\varepsilon) 
\ln (1+\bff(\theta'+i\varepsilon)) 
-\int_{-\infty}^{\infty}d\theta'\ \Gf(\theta-\theta'+i\varepsilon) \ln (1+\bar 
\bff(\theta'-i\varepsilon))\non \\
&+& \int_{-\infty}^{\infty}d\theta'\ \Gf_{2}(\theta-\theta'+i\varepsilon) \ln 
(1+ \yf(\theta'-i\varepsilon)) + i 2mL \sinh \theta + i\,  
\Pf_{bdry}(\theta) -i\pi\,, 
\non \\
\ln \bar \bff(\theta) &=& 
-\int_{-\infty}^{\infty}d\theta'\ \Gf(\theta-\theta'-i\varepsilon) 
\ln (1+\bff(\theta'+i\varepsilon)) 
+\int_{-\infty}^{\infty}d\theta'\ \Gf(\theta-\theta'+i\varepsilon) \ln (1+\bar 
\bff(\theta'-i\varepsilon))\non \\
&+& \int_{-\infty}^{\infty}d\theta'\ \Gf_{2}(\theta'-\theta+i\varepsilon) \ln 
(1+\yf(\theta'+i\varepsilon)) - i 2mL \sinh \theta - i\,  
\Pf_{bdry}(\theta) + i\pi \,, \non \\
\ln \yf(\theta) &=& \int_{-\infty}^{\infty}d\theta'\ 
\Gf_{2}(\theta-\theta'+i\varepsilon) \ln 
(1+\bar \bff(\theta'-i\varepsilon)) + \int_{-\infty}^{\infty}d\theta'\ 
\Gf_{2}(\theta'-\theta+i\varepsilon) \ln 
(1+\bff(\theta'+i\varepsilon))  \non \\
&+& i\, \Pf_{y}(\theta) \,. 
\label{NLIEspin1coord}
\ee
where following definitions have been used,
\be
\varepsilon= \frac{\pi \epsilon}{\mu}\,, \quad \!
\bff(\theta)= b(\frac{\mu \theta}{\pi})\,, \quad \!
\yf(\theta)= y(\frac{\mu \theta}{\pi})
\label{mathfrakdefs}
\ee 
Moreover, $\Gf(\theta)$ and $\Gf_{2}(\theta)$ are taken to be
\be
\Gf(\theta) &=&  \frac{\mu}{\pi} G(\frac{\mu \theta}{\pi}) \non\\
&=& {\mu\over 2\pi^{2}} \int_{-\infty}^{\infty}dk\ e^{-i 
k \mu \theta/\pi}\ \widehat G(k) \,,
\ee
\be
\Gf_{2}(\theta) &=& \frac{\mu}{\pi} G_{2}(\frac{\mu \theta}{\pi}) \non\\
&=&  {\mu\over 2\pi^{2}} \int_{-\infty}^{\infty}dk\ e^{-i 
k \mu \theta/\pi}\ \widehat G_{2}(k) 
= {i\over 2\pi \sinh \theta} \,,
\label{G2theta}
\ee
where 
$\widehat G(k)$ and $\widehat G_{2}(k)$ are as given in (\ref{Gspin1}) and (\ref{G2def}) respectively. Definitions of Fourier transform as given in footnote
4 have been employed. Similarly, $\Pf_{bdry}(\theta)$ and $\Pf_{y}(\theta)$  are given by
\be
\Pf_{bdry}(\theta) &=& P_{bdry}(\frac{\mu \theta}{\pi})\non \\
&=&  {\mu\over 4\pi^{2}} \int_{-\theta}^{\theta} d\theta' \int_{-\infty}^{\infty}dk\ 
e^{-i k \mu \theta'/\pi}\ \hat R(k)\,,
\label{pbdry}
\ee
and
\be
\Pf_{y}(\theta) &=& P_{y}(\frac{\mu \theta}{\pi})\non \\
&=& 4\pi \int_{-\infty}^{\theta}d\theta'\ 
\Gf_{2}(-\theta') = -2i \ln 
\tanh \frac{\theta}{2} - 2\pi \,,
\label{Pytheta}
\ee
respectively. In (\ref{pbdry}), $\hat R(k)$ is given by (\ref{Rk}). As will be shown in subsequent sections, boundary terms $\Pf_{bdry}(\theta)$ and $\Pf_{y}(\theta)$
 are essential when computing the boundary $S$ matrix.

\section{IR limit}\label{sec:IR}

In this section, we consider the IR limit of the NLIEs reviewed in section 3.2, namely the set of equations listed in (\ref{NLIEspin1coord}). 
The IR limit consists of taking $mL\rightarrow\infty$. In this limit, the NLIEs for a one-hole state with rapidity $\theta_{h}$ becomes,
\be
\ln \bff(\theta) &=&  i 2mL \sinh \theta + i \Pf_{bdry}(\theta) 
+ i \varrho(\theta-\theta_{h}) + i \varrho(\theta+\theta_{h}) \non \\
&+& \int_{-\infty}^{\infty}d\theta'\ \Gf_{2}(\theta-\theta'+i\varepsilon) \ln 
(1 + \yf(\theta'-i\varepsilon)) -i\pi \,, \label{nlieb} \\
\ln \yf(\theta) &=& i \Pf_{y}(\theta) + 
i g_{y}(\theta-\theta_{h})
+ i g_{y}(\theta+\theta_{h}) \label{nliey}\,,
\ee
where $\varrho(\theta)$ and $g_{y}(\theta)$ are the hole source terms (see \cite{ArpadRavSuz, Su1})
\be
\varrho(\theta) = 2\pi \int_{0}^{\theta}d\theta'\, \Gf(\theta') \,, 
\qquad 
g_{y}(\theta) = -i \ln \tanh \frac{\theta}{2} + \frac{\pi}{2}
\,. 
\ee 
As pointed out and employed in \cite{ArpadRavSuz, ANS} for cases with periodic and Dirichlet boundary conditions, we stress here that when 
$mL\rightarrow\infty$, the integral terms in (\ref{NLIEspin1coord}) with $\bff(\theta)$ (and $\bar \bff(\theta)$) are of the order $O(e^{-mL})$ and can therefore be neglected. 
The integrals involving $\yf(\theta)$ do not enjoy such property and thus remain after the limit is taken. It is left to be shown in the next section that 
this will lead to a relation equivalent to the Yang equation for a particle on an interval. This will be exploited to compute the boundary $S$ matrix.

\subsection{Boundary $S$ Matrix and lattice-IR relation}\label{sec:bdS}

As for the bulk theory \cite{ABL, ABL2},  in \cite{BPT}, Bajnok {\it et al.} suppose that the full BSSG boundary $S$ matrix is a product of the SG and 
RSOS boundary $S$ matrices.  Also refer to \cite{ANS} where complete boundary $S$ matrix
of this form for the BSSG models with Dirichlet boundary conditions is given. 

In this section, we shall also find such a form here. As pointed out in \cite{ArpadRavSuz, Su1} for the periodic boundary conditions and in \cite{ANS} for 
the case with Dirichlet boundary conditions, here we also have that $\ln \bff(\theta_{h})$ is $i\pi$ times an odd integer which translates into the following 
\be
e^{i 2mL \sinh \theta_{h}}\, e^{i\Pf_{bdry}(\theta_{h}) +i 
\varrho(2\theta_{h})+{\cal L}(\theta_{h})} = 1 \,, \label{nlieexp}
\ee 
which is obtained after (\ref{nlieb}) is evaluated at $\theta_{h}$ followed by exponentiation. In (\ref{nlieexp}), ${\cal L}(\theta)$ is given by
\be
{\cal L}(\theta) \equiv 
\int_{-\infty}^{\infty}d\theta'\ \Gf_{2}(\theta-\theta'+i\varepsilon) 
\ln (1+ \yf(\theta'-i\varepsilon)) \,. \label{calydef}
\ee 
Thus, (\ref{nlieexp}) should be equivalent to the Yang equation for a particle on an interval of length $L$. For an excellent discussion on this matter for the BSG 
model with general integrable boundary interactions, refer to section 4.2 of \cite{ABNPT}. Hence comparison of these two results gives
the following for the product of boundary $S$ matrices,
\be
R(\theta_{h}\,; \lambda\,, \eta_{-}\,, \vartheta_{-})\, 
R(\theta_{h}\,; \lambda\,, \eta_{+}\,, \vartheta_{+})  = e^{i\Pf_{bdry}(\theta_{h}) +i 
\varrho(2\theta_{h})+{\cal L}(\theta_{h})} \,. \label{prodRR}
\ee
Note that $\lambda$ is the IR bulk SSG parameter while $\eta_{\pm}$ and $\vartheta_{\pm}$ are the IR boundary SSG parameters.
We first evaluate the factor $e^{{\cal L}(\theta_{h})}$. The boundary term $ \Pf_{y}(\theta)$ given by (\ref{Pytheta}) is essential in this computation. 
As found in \cite{Murgan3}, this coincides with the expression obtained in \cite{ANS}. In addition, we remark that the term $\Gf_{2}(\theta)$ (and hence its Fourier 
transform, $\widehat G_{2}(k)$) and equation (\ref{nliey}) are identical to the corresponding ones in \cite{ANS} for Dirichlet BSSG${}^{+}$ models. 
Consequently, the computation and the result for the factor $e^{{\cal L}(\theta_{h})}$ are also identical and we therefore choose to omit the steps here. 
We urge the readers to refer to \cite{ArpadRavSuz, ANS} for details. Rather, we present only the result below,
\be
{d\over d\theta_{h}} {\cal L}(\theta_{h}) = \frac{i}{4}\int_{-\infty}^{\infty} dk\, {e^{-2 i k \theta_{h}}\over
\cosh^{2} \frac{\pi k}{2} \cosh^{2} \pi k} \,,
\ee
which after integration and letting $k = \frac{t}{\pi}$ yields, 
\be
e^{{\cal L}(\theta_{h})} \sim P(\theta_{h})^{2} \,,
\label{calLresult}
\ee 
where $P(\theta)$ is given by
\be
P(\theta) \sim \exp \left\{ \frac{i}{8}\int_{0}^{\infty} {dt\over t}\, 
{\sin (2 t \theta/\pi)\over \cosh^{2}\frac{t}{2} \cosh^{2}t}
\right\} \,, 
\label{Pm}
\ee
Indeed (\ref{Pm}) is a reflection factor of the boundary tricritical Ising model \cite{AK}-\cite{NepSup}, the integral representation of which is given in \cite{ANtricitical}. 
This is the RSOS factor. In \cite{BPT}, this factor is given 
in the form of infinite products of gamma functions (Refer to equation (14) of the reference.). 
As before in \cite{ANS}, this result is given only up to crossing factors of the form $e^{const \theta}$.

The remaining term $e^{i\Pf_{bdry}(\theta_{h}) +i 
\varrho(2\theta_{h})}$ in (\ref{prodRR}) will be shown below to represent the SG factor. At this point, we first proceed to compute
${d\over d\theta_{h}}{\Pf_{bdry}}(\theta_{h})$. First, recalling (\ref{pbdry}) and (\ref{Rk}), then differentiating with respect to $\theta_{h}$,
we obtain the following
\be
{d\over d\theta_{h}}\Pf_{bdry}(\theta_{h}) &=& \int_{-\infty}^{\infty} dk\, 
e^{-i k \theta_{h}} \Bigg\{
{\left[s_{+}\sinh\left((\pi - \mu|A_{+}|)\frac{\pi k}{2\mu}\right)+
\sinh\left(\frac{k\pi}{2}B_{+}\right)\right]\over{
2\cosh{\pi k\over{2}}\sinh\left((\frac{\pi}{\mu}-2)\frac{\pi 
k}{2}\right)}}\non \\
 &+& (+ \leftrightarrow -)+ {\cosh \frac{\pi k}{4}\sinh\left((\frac{\pi}{\mu}-3)\frac{\pi k}{4})\right)\over
\cosh \frac{\pi k}{2} \sinh \left((\frac{\pi}{\mu}-2)\frac{\pi k}{4}\right)}
\Bigg\} \,. 
\label{Pfprime}
\ee 
The symbol $(+ \leftrightarrow -)$ represents the terms with $A_{+}\rightarrow A_{-}$, $B_{+}\rightarrow B_{-}$, $s_{+}\rightarrow s_{-}$. 
Similarly the second term gives
\be
{d\over d\theta_{h}}\varrho(2\theta_{h}) &=& \int_{-\infty}^{\infty} dk\, 
e^{-i k \theta_{h}}{\sinh\left((\frac{\pi}{\mu}-3)\frac{\pi k}{4})\right)\over
2\cosh \frac{\pi k}{4} \sinh \left((\frac{\pi}{\mu}-2)\frac{\pi k}{4}\right)} \,. 
\ee 
which together with (\ref{Pfprime}) results in
\be
{d\over d\theta_{h}}\left[\Pf_{bdry}(\theta_{h}) + 
\varrho(2\theta_{h})\right] &=& \int_{-\infty}^{\infty} dk\, 
e^{-i k \theta_{h}} \Bigg\{
{\left[s_{+}\sinh\left((\pi - \mu|A_{+}|)\frac{\pi k}{2\mu}\right)+
\sinh\left(\frac{k\pi}{2}B_{+}\right)\right]\over{
2\cosh{\pi k\over{2}}\sinh\left((\frac{\pi}{\mu}-2)\frac{\pi 
k}{2}\right)}}\non \\
 &+& (+ \leftrightarrow -)+ 2{\sinh \frac{3\pi k}{4}\sinh\left((\frac{\pi}{\mu}-3)\frac{\pi k}{4})\right)\over
\sinh \pi k \sinh \left((\frac{\pi}{\mu}-2)\frac{\pi k}{4}\right)}
\Bigg\} \,, 
\label{SGtermforSSG}
\ee 
Equation (\ref{SGtermforSSG}) resembles the soliton reflection amplitude of the BSG model. Indeed by comparing with 
the soliton reflection amplitude $P_{+}(\theta \,, \eta, \vartheta)$ of the BSG model \cite{GZ},
or more specifically $P_{+}(\theta \,, \eta_{-}, \vartheta_{-})P_{+}(\theta \,, -\eta_{+}, \vartheta_{+})$ for two boundaries, which has the following integral representation, 
\cite{ABNPT} \footnote{See equation (4.35) of this reference.}
\be
\frac{1}{i}\frac{d}{d\theta}\ln \left[P_{+}(\theta \,, \eta_{-}, \vartheta_{-})P_{+}(\theta\,, -\eta_{+}, \vartheta_{+})\right]
&=& \int_{-\infty}^{\infty}dk\, e^{-i k \theta} \Bigg\{
{\sinh\left((1- \frac{2\eta_{+}}{\pi \lambda})\frac{\pi k}{2}\right)\over
2 \cosh\frac{\pi k}{2} \sinh \frac{\pi k}{2\lambda}}
+ {\sinh\left((1+ \frac{2i\vartheta_{+}}{\pi \lambda})\frac{\pi k}{2}\right)\over
2 \cosh\frac{\pi k}{2} \sinh \frac{\pi k}{2\lambda}} \non\\
&+& {\sinh\left((1+ \frac{2\eta_{-}}{\pi \lambda})\frac{\pi k}{2}\right)\over
2 \cosh\frac{\pi k}{2} \sinh \frac{\pi k}{2\lambda}}
+ {\sinh\left((1+ \frac{2i\vartheta_{-}}{\pi \lambda})\frac{\pi k}{2}\right)\over
2 \cosh\frac{\pi k}{2} \sinh \frac{\pi k}{2\lambda}} \non \\
&+& 2{\sinh \frac{3\pi k}{4}\sinh\left((\frac{1}{\lambda}-1)\frac{\pi k}{4})\right)\over
\sinh \pi k \sinh \frac{\pi k}{4\lambda}} \Bigg\}
\label{SGGZ}
\,,
\ee 
one will be able to determine the relation between the boundary lattice and boundary IR parameters. In (\ref{SGGZ}), $\lambda$ is the bulk IR 
parameter,  $\eta_{\pm}$ and $\vartheta_{\pm}$ are the boundary IR parameters. 
Recalling the bulk ``lattice $\leftrightarrow$ IR'' relation (see e.g. \cite{ANS}) for the SSG model,
\be
\lambda = \frac{1}{\frac{\pi}{\mu} - 2}
\label{bulkIR}
\ee
and making the following identification for the boundary ``lattice $\leftrightarrow$ IR'' relation for the BSSG model, 
\be
\eta_{\pm} &=& \mp\frac{\pi}{2} 
\left(s_{\pm}(1+2\lambda) - 2a_{\pm}\lambda\right)\,,
\label{boundarylatIR1}\\
\vartheta_{\pm} &=& \lambda \pi b_{\pm}\,.
\label{boundarylatIR2}
\ee 
one finds the following,
\be
e^{i\Pf_{bdry}(\theta_{h}) +i \varrho(2\theta_{h})} = P_{+}(\theta_{h} \,, 
\eta_{-}\,, \vartheta_{-})\, P_{+}(\theta_{h} \,, -\eta_{+}\,, \vartheta_{+}) \,. \label{SGpart}
\ee 
The above relations between the boundary IR and lattice parameters are among the main results of this paper. Such a relation was 
found in \cite{ANS} for the BSSG models with Dirichlet boundary conditions. Note that (\ref{boundarylatIR2}) resembles the corresponding expression found for
the BSG model in \cite{AN}. However, for the SG model, the bulk ``lattice $\leftrightarrow$ IR'' relation is given by 
$\lambda = \frac{1}{\frac{\pi}{\mu}- 1}$ instead of (\ref{bulkIR}). Furthermore, we can also conclude that, 
combining the results (\ref{prodRR}), (\ref{calLresult}) and (\ref{SGpart}), that the NLIE generates the following SSG boundary $S$ matrices
\be
R(\theta_{h}\,; \lambda\,, \eta_{+}\,, \vartheta_{+}) \sim P_{+}(\theta_{h} \,, 
-\eta_{+}\,, \vartheta_{+}) \, P(\theta_{h}) \non \\
R(\theta_{h}\,; \lambda\,, \eta_{-}\,, \vartheta_{-}) \sim P_{+}(\theta_{h} \,, 
\eta_{-}\,, \vartheta_{-}) \, P(\theta_{h})\,. 
\label{IRresult}
\ee 
which agrees with the boundary $S$ matrix proposed by Bajnok {\it et al.} in \cite{BPT} for the BSSG${}^{+}$ model. This is also our main result. 
As pointed out earlier, the BSSG boundary $S$ matrix is a product of SG and RSOS boundary $S$ matrices.

\section {Discussion}\label{sec:dis}

In this paper, we analyze the IR limit of a set of NLIEs that describe the BSSG model for a state of one hole. 
These NLIEs were obtained previously in \cite{Murgan3} by taking the continuum limit of the NLIEs of the inhomogeneous open spin-$1$ XXZ quantum spin chain 
where the lattice boundary parameters (that form general integrable boundary terms as opposed to diagonal integrable boundary terms treated in \cite{ANS}) satisfying 
a pair of real constraints (\ref{realconstraints}). Thus, these NLIEs should correspond to more general boundary conditions than that considered in \cite{ANS}. 
We computed the boundary $S$ matrix which is the product of RSOS and SG terms. The SG term is then used to propose the relation between the boundary lattice 
and IR parameters.  

One could also further investigate this model. One could investigate the excited states of the model. It should also be possible to carry out such analysis
for open spin-$1$ XXZ chain with non-diagonal boundary terms, the solutions to which are given in \cite{M2}. In contrast to the solution used in this paper, 
the solutions given in \cite{M2} are not restricted by any type of constraints among the lattice boundary parameters. It would be interesting to work out the 
NLIEs for these cases as well and analyze their UV and IR limits. Also, it would be interesting to see if there is an integrable spin-$1$ model that corresponds to
the BSSG$^{-}$ model, explored in \cite{BPT}. We hope to be able to address some of these issues in the future.

\end{document}